\documentclass[twocolumn]{biophys-new}
\usepackage[utf8]{inputenc}
\usepackage{graphicx}
\usepackage[colorlinks,allcolors=cyan!70!black]{hyperref}

\usepackage[version=4]{mhchem}

\usepackage{booktabs,tabularx}

\usepackage{bm}

\usepackage{tikz}
\usetikzlibrary{arrows.meta}

\newcommand\dd{\mathrm{d}}
\newcommand\pp{\partial}

\newcommand\x{\bm{x}}

\renewcommand{\eqref}[1]{\hyperref[#1]{Eq.~\ref*{#1}}}

\usepackage{subcaption}
\usepackage[normalem]{ulem}

\usepackage{tikz}
\usetikzlibrary{arrows.meta,positioning,shapes.geometric}
\definecolor{ucrblue}{RGB}{0,51,102}
\definecolor{softgray}{RGB}{245,246,248}
\definecolor{darkred}{RGB}{145,30,30}
\definecolor{accentyellow}{RGB}{252,243,207}

\usepackage{pgfplots}

\tikzstyle{box} = [rectangle, rounded corners, minimum width=3.2cm, minimum height=1.15cm,
                   text centered, draw=black, fill=blue!10, align=center]
\tikzstyle{arrow} = [thick,->,>=stealth]
\tikzstyle{concept} = [rectangle, dashed, draw=orange!80!black, fill=orange!6,
                       minimum width=3.6cm, minimum height=1.15cm, align=center]

\usepackage{lipsum}

\title{Thermodynamically Consistent Modeling of ATP-Driven Cross-Bridge Dynamics in Muscle Contraction}
\runningtitle{ATP-Driven Cross-Bridge Dynamics} %% For page header

\author[1,*]{Yiwei Wang}
\author[2]{Chun Liu}
\runningauthor{Y. Wang and C. Liu} %% For page header

\affil[1]{Department of Mathematics, University of California, Riverside, Riverside, CA, 92521, United States
}
\affil[2]{Department of Applied Mathematics, Illinois Institute of Technology, Chicago, IL, 60616, United States}

\corrauthor[*]{yiweiw@ucr.edu}

\papertype{Article}
\begin{document}

\begin{frontmatter}

\begin{abstract}
Muscle contraction is a prototypical multiscale chemomechanical process in which ATP hydrolysis at the molecular level drives force generation and mechanical work at larger scales. A central challenge is to incorporate the free energy supplied by ATP hydrolysis into a mechanical cross-bridge model in a way that is thermodynamically consistent and connects microscopic motor cycling to macroscopic force generation.  Here we use the Energetic Variational Approach (EnVarA) to combine Hill's cycle-affinity viewpoint with Huxley's sliding-filament mechanics in a single thermodynamically consistent framework. We formulate a three-state Fokker--Planck--jump description for cross-bridge densities evolving on state-dependent free-energy landscapes, and treat ATP, ADP, and \(\mathrm{P_i}\) as reacting and diffusing chemical species that share the same free-energy functional as the cross-bridge densities. ATP hydrolysis enters the model through local detailed balance, which biases the transition rates. Filament sliding velocity is incorporated as a convective transport term in the Fokker--Planck dynamics, so mechanical power output is defined directly from the energy balance law. Under chemostatted conditions and a fast-equilibration assumption for the two attached states, the model reduces to a closed two-state molecular motor description; in a further small-diffusion limit, this reduction recovers a Huxley-type transport-reaction equation whose effective attachment and detachment rates are \emph{derived} from the underlying three-state cycle rather than prescribed phenomenologically, with their strain, ATP, and phosphate dependence fixed by the energy landscapes and local detailed balance. We calibrate the reduced model against cross-bridge-level observables and show that it reproduces key velocity-dependent trends and yields a Hill-like force--velocity relation comparable to established cross-bridge models.
\end{abstract}

\begin{sigstatement}
We present a thermodynamically consistent variational modeling framework for cross-bridge dynamics that combines a three-state chemomechanical cycle
with Fokker--Planck equations on state-dependent energy landscapes. The full model treats ATP, ADP, and \(\mathrm{P_i}\) as reacting and diffusing species within the same free-energy structure as the cross-bridge densities, extending chemomechanical coupling beyond the chemostatted setting and yielding a closed energy-dissipation balance that tracks chemical input, mechanical power exchange, and irreversible dissipation. Macroscopic filament sliding enters
as a transport term, so external mechanical work is defined through the prescribed free energy rather than imposed separately. Under chemostatted conditions and a fast-equilibration closure, the model reduces to a
three-state, then two-state, FP-jump model, and then a Huxley-type transport--reaction equation whose
effective rates are derived from the three-state cycle. Proof-of-concept simulations reproduce both cross-bridge-level observables and Hill-type force--velocity behavior.
\end{sigstatement}
\end{frontmatter}

\section*{Introduction}

\begin{figure*}[!t]
\centering
\includegraphics[width= 0.9 \textwidth]{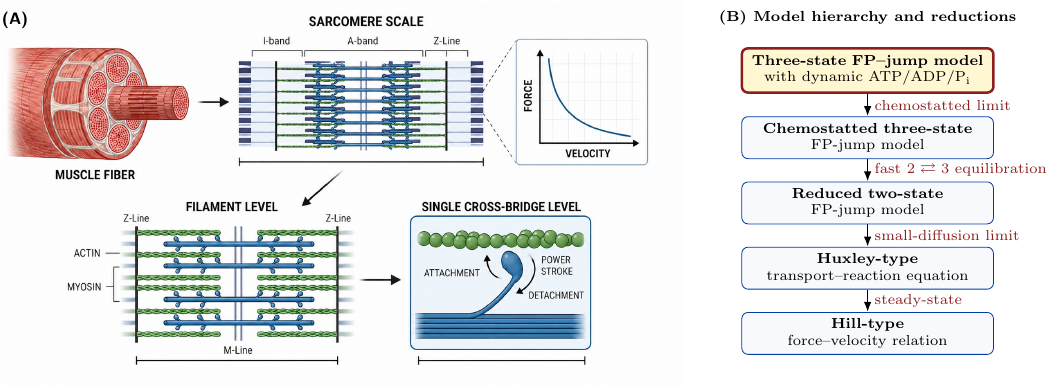}
\caption{Schematic overview of the multiscale muscle structure and the model hierarchy considered in this work. (A) Muscle organization from the fiber and sarcomere scales to the filament and single-cross-bridge levels, together with the
macroscopic force--velocity relation. (B) Formal reduction from the thermodynamically consistent three-state Fokker--Planck--jump model with dynamic nucleotide species to chemostatted and reduced two-state descriptions, and finally to a
Huxley-type transport--reaction equation whose steady state yields a
Hill-type force--velocity relation. Here, fast $2\rightleftarrows3$ equilibration means that the interconversion between the attached pre- and post-stroke states shown in Fig.~\ref{fig:three-state-cycle} occurs on a shorter timescale than transport and overall ATP turnover, allowing the two attached states to be combined into an effective attached state.}\label{fig:multiscale_schematic}
\end{figure*}

Muscle contraction is a prototypical \emph{multiscale} chemomechanical system that underlies animal locomotion and cardiovascular function. Muscle can be considered as an active viscoelastic material \cite{fung2013biomechanics, qian2008viscoelasticity}.
Its function emerges from a complex multiscale chemo-mechanical interaction: ATP hydrolysis and conformational changes of myosin at the nanometer scale generate force and motion at the level of individual cross-bridges \cite{hirokawa1998kinesin}; these microscopic events synchronize within the ordered architecture of sarcomeres and myofibrils, and ultimately give rise to macroscopic mechanical responses of fibers, tissues, and whole organs. Moreover, muscle, particularly \emph{striated} muscle, is one of the few active biological materials for which \emph{multiscale data} are available in a comparatively systematic manner. Structural and dynamical information spanning near-atomic resolution to sarcomere-level organization can be probed by X-ray diffraction and related techniques \cite{powers2021sliding,prodanovic2023using}, while mesoscopic and macroscopic mechanical behavior is routinely quantified through assays such as laser diffractometry and force--velocity measurements \cite{powers2021sliding}. This unique combination of a strongly multiscale mechanism and richly multiscale measurements makes muscle an ideal testbed for developing predictive, thermodynamically consistent models that \emph{bridge} molecular energy transduction and emergent constitutive behavior.

There is a long history of developing mathematical models for muscle contraction.
At the macroscopic scale, the active behavior of muscle can be described by a simple but remarkable force--velocity relation
\begin{equation}\label{Hill_1}
    V = b (F_0 - F)/(F + a)
\end{equation}
proposed by A.V. Hill in 1938 \cite{hill1938heat}. Here, $V$ is the constant shortening velocity of muscle, $F$ is the constant force, $a$, $b$ and $F_0$ are model parameters. One can note that $V_{\rm max} =  \frac{b F_0}{a}$, which is the velocity at which muscle shortens when there is no load. Also, $V = 0$ when $F = F_0$. Hence $F_0$ is called the isometric or stall force. %  which is the force that the muscle will develop 
One can rewrite \eqref{Hill_1} as $b (F_0 - F) = F V + aV,$
which formally corresponds to the first law of thermodynamics $d U = \dd W + \dd Q$, where $a V$ is the heat and $F V$ is the work. The phenomenological parameter $a$ characterizes the frictional heat generated by unloaded muscle shortening, and $b$ is the product of the muscle ATPase rate and the myosin step size   \cite{baker2024mixed}.
%The two parameters, $a$ and $b$, are the frictional heat
%generated by unloaded muscle shortening and the product of the muscle ATPase rate and the myosin step size

% Understanding the force-velocity relation and its molecular basis has been a long standing challenge.
%individual cross-bridge behaviors, 

To explain the physiological origin and molecular mechanism of Hill's equation, at the microscopic scale, A.F. Huxley proposed a simple two-state model to describe the 
myosin--actin interaction \cite{huxley1957muscle}, based on the sliding-filament hypothesis \cite{huxley1954changes, huxley1954structural}.
In Huxley's model, the myosin heads are modeled as spring-like elements that
can attach to and detach from actin with displacement-dependent rates; the distribution of attached cross-bridges evolves under filament sliding and reaction kinetics. Let $x$ denote the relative displacement of an actin
binding site with respect to a reference position of the myosin head, and assume that, at any
instant, each head can interact with at most one nearby binding site. A detached head attaches to the actin site at relative displacement $x$ with a rate $f(x)$. Conversely, due to the free-energy releasing process (ATP hydrolysis), the myosin head can detach from actin with a rate $g(x)$. With $n(x,t)$ being the fraction of attached cross-bridges, the dynamics is described by
\begin{equation}\label{Huxley}
 \frac{\pp n}{\pp t} - V \frac{\pp n}{\pp x} = (1 - n(x))f(x) - g(x) n(x),
 \end{equation}
 where $V > 0$ is the shortening (contracting) velocity. Here, we follow Huxley's original sign convention that shortening velocity carries cross-bridges with a certain displacement into smaller values of displacement \cite{williams2011huxley}. The total force is obtained by multiplying the number of attached cross-bridges by the average force $p$ generated per attached bridge.
%The macroscopic force is obtained by averaging the force generated by attached bridges over the steady attached distribution. 
With suitable velocity-dependent assumptions on $f(x)$, $g(x)$, and $p$, Huxley-type models can reproduce force--velocity measurements \cite{seow2013hill}.
In this formulation, however, the effect of ATP hydrolysis is not made explicit: rather than entering through a thermodynamic affinity $\Delta\mu = \mu_{\rm ATP} - \mu_{\rm ADP} - \mu_{\rm P_i}$ or through the nucleotide concentrations, its role is absorbed into the empirically prescribed rate functions $f(x)$ and $g(x)$. 

The Huxley model has become the foundation for a vast body of work on
cross-bridge kinetics and muscle mechanics
\cite{huxley1971proposed, lymn1971mechanism, peskin1975mathematical,greene1980cooperative,zahalak1981distribution, williams1984,Lacker_peskin_1986,pate1989model, williams1990purely, vilfan1999force, cordova1992dynamics,piazzesi1995cross, smith1995strain,Smith1995,duke1999molecular,chin2006mathematical,smith2008towards,smith2008toward,marcucci2010muscle,seow2013hill,mijailovich2016three,caruel2018physics, regazzoni2021active}.
Broadly speaking, subsequent developments have followed two directions \cite{qian2005cycle,caruel2018physics}. The
first is a \emph{chemistry-based} approach, which refines the original
two-state cross-bridge picture into a finite network of biochemical
states and strain-dependent transition rates, such as the classical
Lymn--Taylor cycle \cite{lymn1971mechanism}. The
second is a \emph{mechanics-based} or \emph{landscape-based} approach,
which describes motor motion as stochastic dynamics on
state-dependent mechanical landscapes, often in the language of
Brownian ratchets or Fokker--Planck equations
\cite{peskin1993cellular, julicher1997modeling, qian2000mathematical,kolomeisky2007molecular}. Although these viewpoints appear different, they can be interpreted as different resolutions of a common
underlying picture: landscape-based stochastic dynamics at
the continuous level (e.g., Klein--Kramers or Fokker--Planck
descriptions) can be coarse-grained, under suitable time-scale
separation, into discrete-state master equations that describe
transitions among metastable states \cite{qian2005cycle}.

A fundamental modeling question is how the concentration of nucleotides affects cross-bridge dynamics and force-velocity relationships \cite{cooke1985effects, chin2006mathematical}. Existing models address this question at several levels. In the original Huxley model, and in many phenomenological multistate cross-bridge variants \cite{seow2013hill}, ATP hydrolysis is not introduced through an explicit thermodynamic affinity
\(
\Delta\mu=\mu_{\rm ATP}-\mu_{\rm ADP}-\mu_{\mathrm{P_i}}.
\)
Instead, its effect is absorbed into empirically prescribed
strain-dependent transition rates. Such models can reproduce
force--velocity curves, but they do not explicitly keep track of how
the chemical free energy of ATP hydrolysis is consumed, converted into
mechanical work, and dissipated.

Classical Fokker--Planck, Brownian-ratchet, and recent jump--diffusion molecular-motor models \cite{peskin1993cellular, julicher1997modeling,parmeggiani1999energy,chaintron2023jump}
incorporate chemical driving more explicitly than phenomenological
Huxley-type rate laws, but are often developed under \emph{chemostatted} conditions: the nucleotide chemical potentials are fixed by external reservoirs, and the effect of ATP hydrolysis enters through a prescribed affinity or chemical potential that biases transitions between state-dependent landscapes, or shifts their relative free-energy levels. However, because the reservoirs are fixed, ATP, ADP, and \(\mathrm{P_i}\) are not evolved as reacting and diffusing chemical fields. Thus ATP-limited regimes, local nucleotide depletion, and supply--demand imbalances are outside the scope of the
chemostatted formulation.

The thermodynamic cycle theory of T.\,L.~Hill and
his collaborators \cite{hill1974theoretical,hill1975some,hill1976theoretical,eisenberg1985muscle} writes the nucleotides directly into the elementary steps of a chemomechanical cycle, with $\Delta\mu$ as the affinity accumulated around a
closed cycle. But Hill's formulation is a discrete-state cycle description: by itself it provides neither the strain-resolved Fokker--Planck transport coupled to filament sliding that Huxley-type mechanics requires, nor reaction--diffusion dynamics for spatially varying nucleotide concentrations. Concrete computational realizations of such cycles, such as Duke's molecular model \cite{duke1999molecular} and the spatially explicit MUSICO framework \cite{mijailovich2016three}, are likewise built on discrete stochastic dynamics rather than on a continuum evolution equation for the cross-bridge probability densities and nucleotide fields.

The purpose of this work is to develop a thermodynamically consistent molecular-motor framework beyond the chemostatted setting, building on T.\,L.~Hill's kinetic-cycle viewpoint while treating ATP, ADP, and \(\mathrm{P_i}\) as dynamical chemical variables. This extension is motivated by the need to describe regimes in which nucleotide supply and turnover are not externally fixed, such as ATP-limited contraction or local nucleotide depletion. We adopt the energetic variational approach (EnVarA) \cite{wang2022some}, which derives a model from a single thermodynamic structure: the available chemical free energy of ATP, ADP, and \(\mathrm{P_i}\) is encoded directly in one free-energy functional through their chemical potentials, rather than appended later as an external driving term. Together with a dissipation functional, this energy-dissipation law generates the Fokker--Planck drift, the jump fluxes satisfying local detailed balance, and the reaction--diffusion dynamics of the nucleotide species. In this formulation, thermodynamic consistency is built into the derivation itself, so chemical driving from ATP hydrolysis, motor transitions, nucleotide source terms, mechanical work, and energy dissipation are all coupled by construction. 

% The classical chemostatted motor model is recovered as a fast-reservoir limitvof this description, in which the nucleotide fields relax instantaneously to prescribed values. Under that chemostatted regime and a fast-equilibration limit between the attached substates, the model reduces further to an effective two-state motor model and, in a singular limit, to a Huxley-type sliding-filament equation.

To connect with classical reduced descriptions, as illustrated in Figure~\ref{fig:multiscale_schematic}, we then consider a
chemostatted regime and a fast-equilibration limit between two attached states, under which the model reduces to an effective two-state
Fokker--Planck--type motor model and, in a further small-diffusion limit, to a
Huxley-type sliding-filament equation, in which the dependence of transition rates $f(x)$ and $g(x)$ on ATP hydrolysis is explicitly derived.
% Although the reduced model is formally of two-state type, it is not the classical two-state scheme: the cycle affinity $\Delta\mu$ of the parent three-state model is inherited by the effective attachment/detachment rates and the effective attached landscape, so the nonequilibrium drive survives the reduction. This both clarifies how chemical driving reshapes the effective rates and landscape, and explains why a genuine three-state cycle is required upstream eventhough the working equations look two-state.

Finally, we provide proof-of-concept numerical experiments for the reduced Huxley-type model. Since the effective attachment and detachment rates  inherit their structure from the underlying thermodynamically consistent
three-state model, we only need to calibrate a few remaining parameters using cross-bridge-level observables in Ref.~\cite{piazzesi2007skeletal}. The resulting reduced model captures key velocity-dependent trends and produces a Hill-type
force--velocity relation comparable to established cross-bridge and multiscale muscle models. These results suggest that the thermodynamically constrained reduced model can serve as a predictive tool for studying how ATP and \(\mathrm{P_i}\)
concentrations, as well as energy-landscape parameters, affect force generation, cross-bridge attachment, mechanical power output, and the resulting force--velocity relation.

\section*{Mathematical Model}

\subsection*{T.L. Hill's cycle thermodynamics}
\label{subsec:hill_duke}

We begin with T. L. Hill's nonequilibrium-thermodynamic viewpoint for cross-bridge cycling \cite{hill2005free}.
In Hill's formulation, mechanical force and chemical kinetics are coupled through local detailed balance.

Let $x$ denote the displacement of the myosin head from its equilibrium position. For state $i$, let $G_i(x)$ be the Gibbs free energy along this coordinate. For mechanically attached states, the force acting on the myosin head due to the potential $G_i(x)$ is $-\partial_x G_i(x)$. Muscle tension, by contrast, is defined as the force exerted by the head against the filament load, and therefore carries the opposite sign. We accordingly define the cross-bridge force generated by each head at state $i$ as $F_i(x) = \partial_x G_i(x)$. Explicit state-dependent landscapes based on an elastic element and a
power-stroke shift were proposed in Duke's stochastic molecular model
\cite{duke1999molecular}.

\begin{figure}[!t]
\centering
\begin{tikzpicture}[
    >=Latex,
    state/.style={
        draw,
        rounded corners,
        align=center,
        font=\scriptsize,
        inner sep=3pt
    },
    rate/.style={
        font=\scriptsize,
        fill=white,
        inner sep=1pt
    },
    every path/.style={line width=0.5pt}
]
\node[state] (s1) at (0,2.0)
    {\textbf{State 1}\\
     \ce{M.ADP.P_i}\\
     detached};

\node[state] (s2) at (-2.5,0)
    {\textbf{State 2}\\
     \ce{A.M.ADP.P_i}\\
     attached, pre-stroke};

\node[state] (s3) at (2.5,0)
    {\textbf{State 3}\\
     \ce{A.M.ADP}\\
     attached, post-stroke};

% State 1 <-> State 2
\draw[->,bend right=10]
    (s1.south west) to
    node[rate,left] {$\alpha_{12}(x)$} (s2.north);
\draw[->,bend right=10]
    (s2.north east) to
    node[rate,left] {$\alpha_{21}(x)$} (s1.south);

% State 2 <-> State 3
\draw[->,bend left=10]
    (s2.east) to
    node[rate,above] {$\alpha_{23}(x)$} (s3.west);
\draw[->,bend left=10]
    (s3.west) to
    node[rate,below] {$\alpha_{32}(x)c_{\mathrm{Pi}}$} (s2.east);

% State 3 <-> State 1
\draw[->,bend right=10]
    (s3.north) to
    node[rate,right] {$\alpha_{31}(x)c_{\mathrm{ATP}}$} (s1.south east);
\draw[->,bend right=10]
    (s1.south) to
    node[rate, pos=0.58, xshift=3mm, yshift=1mm] {$\alpha_{13}(x)c_{\mathrm{ADP}}$} (s3.north west);

\end{tikzpicture}
\caption{Three-state chemomechanical cycle for a single cross-bridge.
Here $\alpha_{ij}(x)$ is the elementary rate coefficient
for the transition $i\to j$. }
\label{fig:three-state-cycle}
\end{figure}

ATP hydrolysis $\ce{ATP <=> ADP + P_i}$ provides the nonequilibrium driving, which we represent by the chemical potential difference
\begin{equation}
\label{eq:dmu}
\Delta\mu = \mu_{\rm ATP}-\mu_{\rm ADP}-\mu_{P_i}.
\end{equation}
In the chemostatted setting, $\Delta\mu$ is treated as a prescribed parameter controlling how far the system is driven from equilibrium. To incorporate the chemical potential difference $\Delta\mu$, Hill introduced
the three-state cross-bridge cycle shown in
Fig.~\ref{fig:three-state-cycle}: state 1,
$\ce{M.ADP.P_i}$, is detached; state 2,
$\ce{A.M.ADP.P_i}$, is attached before the power stroke; and state 3,
$\ce{A.M.ADP}$, is attached after the power stroke. The three elementary
reversible reactions are $\ce{M.ADP.P_i <=> A.M.ADP.P_i}$,
$\ce{A.M.ADP.P_i <=> A.M.ADP + P_i}$, and
$\ce{A.M.ADP + ATP <=> M.ADP.P_i + ADP}$. Let $\alpha_{ij}(x)$ denote the elementary rate coefficient for the transition
$i\to j$ at displacement $x$. The concentration of any participating soluble
nucleotide enters the corresponding transition rate as indicated in the
figure. Accordingly, we define the transition rates
$w_{12}=\alpha_{12}(x)$, $w_{21}=\alpha_{21}(x)$,
$w_{23}=\alpha_{23}(x)$, $w_{32}=\alpha_{32}(x)c_{P_i}$,
$w_{31}=\alpha_{31}(x)c_{\rm ATP}$, and
$w_{13}=\alpha_{13}(x)c_{\rm ADP}$.
Thermodynamic consistency is enforced by local detailed balance,
\begin{equation}
\label{eq:ldb_general}
\frac{w_{ij}(x)}{w_{ji}(x)}
=
\exp\!\left(
-\beta\,[G_j(x)-G_i(x)+\Delta\mu_{ij}]
\right),~~\beta=(k_BT)^{-1},
\end{equation}
where $\Delta\mu_{ij}$ accounts for the chemical potential differences
associated with the elementary step; in particular,
$\Delta\mu_{12}=0$, $\Delta\mu_{23}=\mu_{P_i}$, and
$\Delta\mu_{31}=\mu_{\rm ADP}-\mu_{\rm ATP}$.
Summing the logarithmic rate ratios around the directed cycle yields the
total affinity $\beta\Delta\mu$:
\begin{equation}\label{eq:hill_cycle_affinity}
\sum_{\text{cycle}}
\log\frac{w_{ij}}{w_{ji}}
=
\frac{\Delta\mu}{k_BT}.
\end{equation}
\eqref{eq:hill_cycle_affinity} characterizes the fundamental
nonequilibrium nature of the system \cite{qian2005cycle}. In a physiological context, the cell's metabolic machinery maintains ATP, ADP, and $P_i$ at concentrations far from their equilibrium ratios \cite{qian2005cycle}. This cycle affinity acts as a non-conservative ``chemical battery'' that breaks detailed balance, sustaining a non-zero net flux through the state network. This driving force prevents the system from relaxing to thermal equilibrium, instead supporting a non-equilibrium steady state (NESS). Conversely, the equilibrium limit $\Delta\mu=0$ restores detailed balance, forcing the steady-state cycle flux to vanish and halting the directed mechanical output \cite{qian2005cycle}.

\iffalse
The thermodynamic affinity $\Delta\mu$ provides the necessary "fuel" for the motor, but the actual transduction of chemical energy into directed mechanical work is governed by the spatial profiles of the free-energy landscapes. In \cite{duke1999molecular}, \citeauthor{duke1999molecular} proposed explicit forms of the free-energy landscapes based on an elastic element and a power-stroke shift, in which
% In a common nondimensional convention $U_i=G_i/(k_BT)$,
\begin{equation}
\label{eq:duke_G}
\begin{aligned}
& G_1(x) = 0, ~~G_2(x) = -\Delta G_{\rm bind} + \frac{\kappa}{2}\,x^2,\\
& G_3(x) = -\Delta G_{\rm bind}-\Delta G_{\rm stroke} + \frac{\kappa}{2}\,(x+d)^2,
\end{aligned}
\end{equation}
where $\kappa$ is an effective stiffness, $d$ is the stroke distance, and $\Delta G_{\rm bind}$ and $\Delta G_{\rm stroke}$ encode binding and power-stroke energy changes.
\fi
% This choice is not a claim of biochemical completeness; it serves as a minimal energetic scaffold on which we build a variationally consistent continuum model.

Although Hill's cycle theory provides the thermodynamic foundation for
chemomechanical coupling, several modeling ingredients must be combined in
order to obtain a continuum description of ATP-driven cross-bridge dynamics.
A strain-resolved Fokker--Planck--jump formulation is needed to replace
individual stochastic motor trajectories by probability densities evolving
on state-dependent energy landscapes, while still retaining microscopic
information such as attached distributions, transition fluxes, and force
generation. At the same time, to go beyond the chemostatted setting, ATP,
ADP, and \(\mathrm{P_i}\) should be treated as reacting and diffusing
chemical variables rather than as fixed external reservoirs. %This is essential for describing regimes in which nucleotide supply, turnover, or local depletion evolve together with the cross-bridge cycle.

The main challenge is that these components cannot be coupled arbitrarily.
The drift on the mechanical landscapes, the jump fluxes between chemical
states, the reaction affinities satisfying local detailed balance, the
transport of nucleotide species, and the mechanical power exchanged with
filament sliding must all be thermodynamically consistent. In particular, the chemical free energy of ATP hydrolysis
should not be appended as an external driving term after the mechanical model has been specified; it should enter the same free-energy structure that determines the cross-bridge dynamics and the nucleotide evolution.

To achieve this, we formulate the model using the energetic
variational approach (EnVarA), which describes a complex system in
terms of its free energy and dissipation mechanisms
\cite{wang2020field,wang2025energetic}. In the present setting, the
cross-bridge probability densities and nucleotide species are governed
by a common free-energy functional, while mechanical diffusion,
chemical transport, and reaction kinetics are represented by distinct
dissipation mechanisms. Within this framework, the transport terms,
reaction laws, and micro--macro power-transfer relations are derived
from the same energy--dissipation structure. The resulting FP--jump
model is therefore thermodynamically consistent by construction, rather
than assembled piece by piece. Further details of the EnVarA derivation
are provided in the supporting material.

%========================
% Setting + Energy + Dissipation (LaTeX-ready)
%========================

\subsection*{A three-state model by EnVarA}
In this subsection, we first derive a thermodynamically consistent three-state model for the case $V(t) = 0$, in which the system is mechanically closed and can be described by a single energy-dissipation law.
Following the classic frameworks of Huxley \cite{huxley1957muscle} and Duke
\cite{duke1999molecular}, we consider a representative myosin head on the
thick filament interacting with at most one nearby binding site on the thin
filament. The continuous internal coordinate $x\in[-l,l]$ denotes their
relative displacement; when the head is attached, $x$ equivalently
describes the position of the bound actin site relative to the equilibrium
position of the head \cite{huxley1954changes}. $l$ can be understood as the finite displacement range accessible to a cross-bridge due to the limited reach of the myosin
head \cite{williams2011huxley}. Since our primary goal is to illustrate the modeling framework rather than to perform a comprehensive physiological description, we intentionally neglect higher-level biological details such as lattice geometry, explicit S1/S2 microstructure and filament compliance beyond an effective internal coordinate, and cooperative interactions among neighboring heads \cite{mijailovich2016three}. More detailed $N$-head formulations that incorporate cooperative effects can be built within the same framework by adopting the setting in \cite{mouri2008fokker}.

Following T. L. Hill's work \cite{hill2005free}, we assume the cross-bridge has three chemical states: state~1 (detached), state~2 (attached pre-stroke),
and state~3 (attached post-stroke), defined in Fig. \ref{fig:three-state-cycle}.
For each $i\in\{1,2,3\}$, let $p_i(x,t)$ be the probability density of being in state~$i$ at $x$ at time $t$.
Thus $p_i$ has units of (length)$^{-1}$ and we impose the normalization $\sum_{i=1}^3\int_{-l}^l p_i(x,t)\,dx=1.$
Let $G_i(x)$ denote the effective free-energy landscape for state~$i$, then the mechanical force generated by the attached states at $x$ is $F_i(x)= \partial_x G_i(x)$ for  $i=2,3.$ Let $c_{\mathrm{ATP}}, c_{\mathrm{ADP}}, c_{P_i}$ denote the molecular line densities (number of molecules per unit length) of ATP, ADP, $\mathrm{P_i}$ in the local periodic micro-environment, and write $\mathcal C:=\{\mathrm{ATP},\mathrm{ADP},P_i\}$ for the corresponding index set, so that $c_j$ denotes the line density of species $j\in\mathcal C$. The three elementary reactions in Hill's cycle are represented by reaction trajectories, i.e., the extents of reaction \cite{oster1974chemical, kondepudi2014modern}, denoted by
$R_{12}(x,t)$, $R_{23}(x,t)$, and $R_{31}(x,t)$.
Their time derivatives, $r_{ab}(x,t):=\dot R_{ab}(x,t)$ for $(ab)\in\{12,23,31\}$, are the reaction rates. For each state $i$, denote by $u_i(x,t)$ the drift velocity in the internal coordinate for heads in state~$i$.
For each chemical species $j\in\mathcal C$, denote by $v_j(x,t)$ the transport velocity
of $c_j$. By mass conservation, $p_i$ ($i=1,2,3$) and $c_j$ ($j\in\mathcal C$) satisfy the kinematics
\begin{equation}\label{eq:kinematics_pi_cj}
 \pp_t p_i + \pp_x (p_i u_i) = J_i, \quad \pp_t c_j + \pp_x (c_j v_j) = S_j \,
\end{equation}
where the source terms are induced by the cycle rates $r_{ab}$:
\begin{equation}\label{eq:sources_pi_r}
J_1=-r_{12}+r_{31},\qquad
J_2=r_{12}-r_{23},\qquad
J_3=r_{23}-r_{31},
\end{equation}
and the chemical production/consumption terms are $S_{\mathrm{ATP}}=-r_{31}, S_{\mathrm{ADP}}=r_{31},$ and $S_{P_i}=r_{23}$.

We define the total free energy on $[-l, l]$ as
\begin{equation}\label{eq:free_energy_RpC}
\begin{aligned}
\mathcal F[\bm p,{\bm c}]
& =\int_{-l}^l\sum_{i=1}^3
\Bigl(k_BT\,p_i \left( \ln\!\frac{p_i}{p^{\rm ref}} - 1 \right) + p_i\,G_i(x)\Bigr)\,dx \\
& +\int_{-l}^l\sum_{j \in\mathcal C}
\Bigl(k_B T \, c_j\left(\ln\!\frac{ c_j}{ c^\circ}-1\Bigr)+ c_j \mu_j^\circ\, \right)\,dx,
\end{aligned}
\end{equation}
where $p^{\rm ref}=1/(2l)$ is a reference density, $c^\circ$ is the reference molecular line density, and $\mu_j^\circ$, $j\in\mathcal C$, are standard chemical potentials. In the free energy, the first part accounts for the configurational entropy and mechanical energy of the cross-bridges, denoted by $\mathcal{F}_{\rm mech}({\bm p})$, while the second part accounts for chemical energy associated with the nucleotide species, denoted by $\mathcal{F}_{\rm chem}({\bm c})$.
The associated chemical potentials are
\begin{equation}\label{eq:chempot_def_RpC}
\begin{aligned}
& \mu_i:=\frac{\delta\mathcal F}{\delta p_i}
= k_BT\ln\!\frac{p_i}{p^{\rm ref}} + G_i(x), \\
& \mu_{c_j}:=\frac{\delta\mathcal F}{\delta c_j}
= k_B T \ln\!\frac{c_j}{ c^\circ}+\mu_j^\circ, \quad j\in\mathcal C. \\
\end{aligned}
\end{equation}

Throughout this section, we impose no-flux boundary conditions on
the cross-bridge probability fluxes at $x=\pm l$, while the chemical
variables $c_j$, $j\in\mathcal C$, are periodic on the representative
chemical domain. These boundary conditions eliminate boundary energy
fluxes and make the representative system energetically closed. In this
setting, $\mathcal F[\bm p,\bm c]$ is a Lyapunov functional: in the
absence of external metabolic input, the initially stored chemical free
energy $\mathcal{F}_{\rm chem} [{\bm c}]$ is dissipated or converted into mechanical work.
Living muscle, however, operates as an open biochemical system.
Cellular metabolism maintains ATP, ADP, and $\mathrm{P_i}$ away from
equilibrium by replenishing ATP and recycling or removing the hydrolysis
products. Once the closed-system equations have been derived, these
metabolic effects can be incorporated through source--sink terms in the
nucleotide equations or, when appropriate, through boundary conditions
representing exchange with the surrounding biochemical environment.

To construct a thermodynamically consistent dynamics via EnVarA, we specify a dissipation rate 
\begin{equation}\label{eq:dissipation_total_r}
\Delta
=\int_{-l}^l\Bigg[
\sum_{i=1}^3 \zeta_i\, p_i\, |u_i|^2
+\sum_{j\in\mathcal C}\varepsilon_j\, c_j\, |v_j|^2
+\Delta_{\rm rxn}
\Bigg]\,dx,
\end{equation}
where $\zeta_i>0$ and $\varepsilon_j>0$, $j\in\mathcal C$, are friction coefficients.
A convenient choice of reaction dissipation that guarantees nonnegative entropy production is \cite{wang2020field}
\begin{equation}\label{eq:dissipation_rxn_r}
\Delta_{\rm rxn}({\bm p}, {\bm c}, \dot{R}_{ab})
=\sum_{(ab) \in \mathcal{I}} k_BT\, \dot{R}_{ab}\,
\ln\!\left(1+\frac{\dot{R}_{ab}}{r_{ab}^-({\bm p}, {\bm c})}\right) \geq 0,
\end{equation}
where $\mathcal{I} =\{ 12, 23, 31 \}$, $r_{ab}^-({\bm p}, {\bm c})>0$ denotes the backward reaction rate for the elementary step $a\to b$.

By the standard variational procedure, the mechanical force balance gives
\begin{equation}\label{eq:flux_closure_main_simple}
\begin{aligned}
& \zeta_i p_i u_i  = -  p_i \partial_x\mu_i,\qquad
\varepsilon_j c_j v_j =  - c_j \partial_x\mu_{c_j}. 
\end{aligned}
\end{equation}
Combining \eqref{eq:flux_closure_main_simple} with the kinematics \eqref{eq:kinematics_pi_cj}, we have
\begin{equation}\label{eq:fp_np_main_simple}
  \begin{aligned}
& \partial_t p_i 
=\partial_x\!\left(\tfrac{k_BT}{\zeta_i}\partial_x p_i + \tfrac{1}{\zeta_i}p_i\,\partial_x G_i(x)\right)+J_i, \\
& \partial_t c_j=\partial_x\!\left(\tfrac{k_B T}{\varepsilon_j}\partial_x c_j\right)+S_j, \quad j\in\mathcal C \\
  \end{aligned}
\end{equation}
For the reaction part, introduce the local affinities \cite{kondepudi2014modern}
\begin{equation}\label{eq:affinities_main}
\begin{aligned}
& \mathcal A_{12}:=\mu_1-\mu_2,\qquad \mathcal A_{23}:=\mu_2-\mu_3-\mu_{c_{P_i}}, \\
& \mathcal A_{31}:=\mu_3-\mu_1-\mu_{c_{\mathrm{ADP}}}+\mu_{c_{\mathrm{ATP}}}. \\
\end{aligned}
\end{equation}
The dissipation $\Delta_{\rm rxn}$ implies
\begin{equation}\label{eq:rxn_closure_main}
\dot R_{ab}
= r_{ab}^{-}(x,t)\Big(\exp\!\big(\mathcal A_{ab}/(k_BT)\big)-1\Big),
\end{equation}
where $ (ab)\in\{12,23,31\}$. By taking the backward rates as
\[
r_{12}^-=\alpha_{21}(x)p_2,~~ r_{23}^-=\alpha_{32}(x)\,c_{P_i}\,p_3,~~ r_{31}^- = \alpha_{13}(x)\,c_{\mathrm{ADP}}\,p_1.
\]
Combining \eqref{eq:affinities_main} and \eqref{eq:rxn_closure_main}
gives the law of mass action:
\begin{equation}\label{eq:mass_action_main}
\begin{aligned}
& \dot R_{12}= \alpha_{12} p_1 - \alpha_{21} p_2,~~\dot R_{23}= \alpha_{23} p_2 - \alpha_{32} c_{P_i}p_3, \\
& \dot R_{31}= \alpha_{31} c_{\mathrm{ATP}}p_3 - \alpha_{13} c_{\mathrm{ADP}}p_1, \\
\end{aligned}
\end{equation}
where the ratios $\alpha_{ab}/\alpha_{ba}$ are determined by the energetic landscapes and the standard chemical potentials through the local detailed balance condition.

\subsection*{Mechanical power output and efficiency}
\label{subsec:efficiency}
The model derived above is a closed chemomechanical system, in which the total free energy decreases monotonically. We now open the system mechanically by prescribing a filament sliding velocity \(V(t)\). Following Huxley's sign convention, filament sliding carries cross-bridges with a given displacement toward smaller displacement values \cite{williams2011huxley}; this generates an additional transport term in the microscopic Fokker--Planck equation:
\begin{equation}\label{FP_three}
 \partial_t p_i -  V \partial_x p_i
=\partial_x\!\left(\tfrac{k_BT}{\zeta_i}\partial_x p_i + \tfrac{1}{\zeta_i}p_i\,\partial_x G_i(x)\right)+J_i.
\end{equation}
Differentiating the free energy (\eqref{eq:free_energy_RpC}) in time and using the equations of $p_i$ and $c_j$ gives
\begin{equation}\label{energy_law_by_eq}
\begin{aligned}
 & \frac{\dd}{\dd t} \mathcal{F}({\bm p}, {\bm c}) = \int_{-l}^l \sum_{i=1}^3 \mu_i \,\pp_t p_i + \sum_{j\in\mathcal C} \mu_{c_j} \,\pp_t c_j\,\dd x \\
  & = - V(t)  \int_{-l}^l  \sum_{i=1}^3 p_i \,\pp_x \mu_i \,\dd x - \Delta_{\rm mech} - \Delta_{\rm c} - \Delta_{\rm rxn}\ ,
\end{aligned}
\end{equation}
under the given boundary conditions of $p_i$ and $c_j$.
Here, $\Delta_{\rm mech} =\int_{-l}^{l}\sum_{i=1}^3 \frac{p_i}{\zeta_i}\,\bigl|\partial_x\mu_i\bigr|^2\,dx$
is the mechanical dissipation for the drift and diffusion of $p_i$,
$\Delta_{\rm c} =\int_{-l}^{l}\sum_{j\in\mathcal C}\frac{c_j}{\varepsilon_j} |\partial_x\mu_{c_j} |^2\,dx$ is the dissipation for diffusion of chemical species, and $\Delta_{\rm rxn}$ is the reaction dissipation, defined as in \eqref{eq:dissipation_rxn_r}. 

The new transport term $-V \pp_x p_i$ contributes the first term on the right-hand side of \eqref{energy_law_by_eq}, which represents the mechanical power transferred from the microscopic cross-bridge dynamics to the macroscopic sliding motion. We define this power as
\begin{equation}\label{eq:P_slide_def}
P_{\rm slide} = V(t) \textstyle \int_{-l}^l  \sum_{i=1}^3 p_i \pp_x \mu_i \dd \x =  V(t) F_{cb} ,
\end{equation}
where 
\begin{equation}\label{def_cb}
F_{\rm cb}(t):=   \sum_{i=1}^3\int_{-l}^{l} p_i(x,t) \pp_x \mu_i \,dx =   \sum_{i=1}^3\int_{-l}^{l} p_i(x,t) \pp_x G_i \,dx,
\end{equation}
is the force generated by the cross-bridges. The second equality holds because, with $\mu_i = k_BT\ln\!\big(p_i/p_{\rm ref}\big)+G_i(x)$, the entropic part of $\mu_i$ integrates to zero when $p_i(\pm l) = 0$.

In \eqref{energy_law_by_eq},
$\Delta_{\rm rxn}\geq0$ is the irreversible dissipation associated with
the reactions in Hill's cycle. To track the chemical free energy released by ATP hydrolysis, we compute its rate of change, $\frac{\dd}{\dd t}\mathcal F_{\rm chem}({\bm c}) = -\Delta_{\rm c}- P_{\rm chem},$
where
\begin{equation}\label{eq:chemical_power_general}
P_{\rm chem}
=
\textstyle \int_{-l}^{l}
[
r_{31}
\bigl(\mu_{\rm ATP}-\mu_{\rm ADP}\bigr)
-
r_{23} \mu_{\mathrm{P_i}}
]\dd x
\end{equation}
is the rate at which chemical free energy is released by the nucleotide reactions. Correspondingly, the mechanical free energy changes as $\frac{\dd}{\dd t}\mathcal F_{\rm mech}({\bm p}) = P_{\rm chem} - P_{\rm slide} - \Delta_{\rm mech}-\Delta_{\rm rxn}.$ We can therefore view $P_{\rm chem}$ as the chemical input to the mechanical subsystem, which is either converted into useful sliding work $P_{\rm slide}$ or dissipated through $\Delta_{\rm mech}+\Delta_{\rm rxn}$. This motivates defining the mechanical efficiency as the ratio of $P_{\rm slide}$ to $P_{\rm chem}$.
%\begin{equation}
%\label{eq:efficiency_def}
%\eta_{\rm ext}(t)=\frac{P_{\rm slide}(t)}{P_{\rm ATP}(t)}
%=  \frac{F_{\rm cb}(t)\,V(t)}{\textstyle \int_{-l}^{l} \dot{R}_{31}(x,t)\,\Delta\mu(x,t)\,dx } .
%\end{equation}
This efficiency measures the fraction of the ATP free energy that
appears as useful external sliding work under a prescribed velocity. Under isometric conditions, $V=0$ and hence $P_{\rm slide}=VF_{\rm cb}=0$, so $\eta=0$. However, $P_{\rm chem}$ need not vanish at $V=0$, since the cross-bridge cycle can continue to turn over ATP and generate force without producing net filament sliding. In this case the entire chemical power is dissipated through $\Delta_{\rm mech}$ and $\Delta_{\rm rxn}$ rather than converted into external work.

\subsection*{Model reduction}

\noindent {\bf Chemostatted regime}. In living muscle, ATP, ADP, and \(\mathrm{P_i}\) are maintained away from equilibrium by metabolic processes \cite{qian2005cycle}, which makes the system \emph{chemically open}.
If the details of this metabolic maintenance are not resolved explicitly, their effect can be represented by prescribing the nucleotide concentrations \(c_{\mathrm{ATP}},c_{\mathrm{ADP}},c_{\mathrm{P_i}}\), or equivalently the chemical potentials \(\mu_{\mathrm{ATP}},\mu_{\mathrm{ADP}},\mu_{\mathrm{P_i}}\). This is known as the chemostatted regime. The chemostatted description is
valid when the nucleotide-reservoir relaxation time $\tau_{\rm res}$ and the
nucleotide diffusion time are both much shorter than the cross-bridge cycle
time, so that $c_{\rm ATP},c_{\rm ADP},c_{\rm P_i}$ can be treated as prescribed, spatially uniform constants. Under the chemostatted condition, the full coupled system therefore reduces to the three-state Fokker--Planck--jump model \eqref{FP_three} for $p_i$ \cite{qian2000mathematical}. A key point of this reduction is that we do not collapse the chemical driving
into a single phenomenological parameter \(\Delta\mu\) that simply tilts the mechanical landscape or modifies the rates globally. Instead, the
fixed ATP, ADP, and $\mathrm{P_i}$ concentrations remain explicit in the
individual reaction rates (\eqref{eq:mass_action_main}).

Under the chemostatted condition, if we assume that $p_i$ equilibrates rapidly and evaluate the efficiency at the steady state of $p_i$, then $P_{\rm chem}$, defined in \eqref{eq:chemical_power_general}, reduces to the chemical
energy consumption rate associated with ATP turnover,
\begin{equation}\label{eq:Pchem_steady}
    P_{\rm ATP}
    :=
    \Delta\mu \textstyle \int_{-l}^{l}r_{31}(x)\,\dd x.
\end{equation}
Indeed, at steady state, $\partial_t p_i=0$; integrating the state-balance equations
and using the no-flux boundary conditions gives $\int_{-l}^{l}J_i(x)\,\dd x=0,$ and hence $\int_{-l}^{l}r_{12}(x)\,\dd x = \int_{-l}^{l}r_{23}(x)\,\dd x = \int_{-l}^{l}r_{31}(x)\,\dd x.$ The mechanical efficiency can therefore equivalently be computed as $\eta=
   P_{\rm slide} / P_{\rm ATP}.$

\noindent {\bf Fast $2 \rightleftarrows 3$ equilibration}. From the three-state model, we can obtain the classical two-state model \cite{julicher1998molecular, qian2000mathematical}  by grouping two attached states, and define
\[
p_-(x,t):=p_1(x,t),\qquad p_+(x,t):=p_2(x,t)+p_3(x,t).
\]
To be more precise, let the drift--diffusion operators be
\[
\mathcal L_i p := \partial_x\!\left(D_i \partial_x p+\tfrac{1}{\zeta_i}p\,\partial_x G_i(x)\right),
\qquad D_i:=\tfrac{k_BT}{\zeta_i}.
\]
Summing the equations for $p_2$ and $p_3$ yields the exact identity
\begin{equation}\label{eq:group_exact}
\partial_t p_+ - V\partial_x p_+
= \mathcal L_2p_2+\mathcal L_3p_3+r_{12}-r_{31},
\end{equation}
which is not closed in terms of $p_+$. In the forward-cycle-dominated regime,
we neglect the reverse transitions $2\to1$ and $1\to3$, so that
$r_{12}\approx\alpha_{12}p_1$ and
$r_{31}\approx\alpha_{31}c_{\rm ATP}p_3$. To close this system, we adopt a fast equilibrium approximation \cite{zahalak2000two}. Specifically, we assume that transition between the attached states $2$ and $3$ is much faster than both the transport timescale and the overall ATP turnover. In other words, $\alpha_{23}$ and $\alpha_{32}$ are of order
$\mathcal O(\epsilon_{23}^{-1})$, while the other rate constants are
of order $\mathcal O(1)$, where $\epsilon_{23}\ll1$ is the small
parameter. Then the (local) detailed-balance relation for the $2 \ce{<=>} 3$ transition implies that
\begin{equation}\label{eq:p2p3_alpha}
p_2 \approx \alpha(x)\,p_+,\qquad p_3 \approx (1-\alpha(x))\,p_+,
\end{equation}
where 
\begin{equation}
\alpha(x) = (1 + \exp(\beta(G_2(x) - G_3(x) - \mu_{P_i})))^{-1}\ ,
\end{equation}
which is determined by the energy landscapes as well as the concentration of ${\rm Pi}$. By further assuming that $\zeta_2 = \zeta_3 = \zeta$ and denoting
$D=k_BT/\zeta$, we have
\begin{equation}
\mathcal{L} p_{+} = \mathcal L_2 p_2+\mathcal L_3 p_3  = \pp_x \left( D \pp_x p_+ + \tfrac{1}{\zeta} p_+ \pp_x G_{+} \right)\ ,
\end{equation}
where \begin{equation}\label{effect_G}
\pp_x G_{+} = \alpha(x) \pp_x G_2 + (1-\alpha(x)) \pp_x G_3.
\end{equation}
One can view $G_{+}$ as the \emph{effective attached free-energy landscape}. Although the three-state model introduces chemical driving through concentration-dependent reaction rates, the reduced description introduces the same driving by reshaping the effective attached landscape $G_{+}$
through $\alpha(x)$. Finally, $(p_-,p_+)$ satisfy the closed two-state coupled-diffusion system
\begin{equation}\label{eq:two_state_closed}
\begin{aligned}
\partial_t p_- - V \partial_x p_- &=  \mathcal{L}_1 p_- - f(x) p_-  +  g(x) p_+,\\
\partial_t p_+ - V \partial_x p_+ &=  \mathcal{L} p_{+} + f(x) p_- - g(x)\,p_+,
\end{aligned}
\end{equation}
which is the classical two-state Fokker-Planck type model for molecular motors \cite{julicher1997modeling, julicher1998molecular, qian2000mathematical, chipot2003variational}.
Here, the effective transition rates are
\begin{equation}\label{eq:fg_effective}
f(x):= \alpha_{12}(x),~~ g(x) = \alpha_{31}(x) c_{\rm ATP} (1-\alpha(x)).
\end{equation}
A direct calculation shows that
\begin{equation}
 \frac{f(x)}{g(x)} = \frac{\alpha_{12}(1 + \exp(-\beta(G_2 - G_3 -\mu_{Pi})))}{\alpha_{31} c_{ATP}}.
\end{equation}

\noindent {\bf Small-diffusion limit}. Finally, in the small-diffusion limit $D_1,D\to0$, equivalently
$\zeta_1^{-1},\zeta^{-1}\to0$, the total cross-bridge density
$\rho=p_-+p_+$ satisfies $\partial_t\rho-V\partial_x\rho=0$.
For a uniform total density $\rho=\rho_0$, defining the attached fraction
$n(x,t)=p_+(x,t)/\rho_0$, the two-state Fokker--Planck system reduces to
the Huxley transport--reaction equation \cite{qian2000mathematical}.

Compared with the original Huxley formulation, the present derivation provides an explicit, thermodynamically consistent characterization of how the attachment/detachment kinetics depend on both mechanical effects (through $G_2,G_3$, and hence the $x$-dependence) and chemical driving (through $\Delta\mu$ via local detailed balance). 
Moreover, it clarifies how the nonequilibrium drive is inherited in the reduced description: $\Delta\mu$ influences the effective attached dynamics through $\alpha(x)$ (and thus the effective landscape $G_+$), while simultaneously biasing the effective reaction kinetics through \eqref{eq:fg_effective}.

\section*{Results for the reduced model}
In this section, we examine whether the reduced Huxley-type model, derived from the thermodynamically consistent three-state model, can reproduce
classical force--velocity behavior, as well as the cross-bridge-level trends
reported in \cite{piazzesi2007skeletal}. The purpose of the numerical experiment is
not to perform a systematic parameter-inference study, but to provide a
proof-of-concept calibration showing that the reduced model is compatible with several experimentally motivated observables, not only
with the overall Hill-type force--velocity curve.

\begin{figure*}[!t]
  \centering
\includegraphics[width=0.9\textwidth]{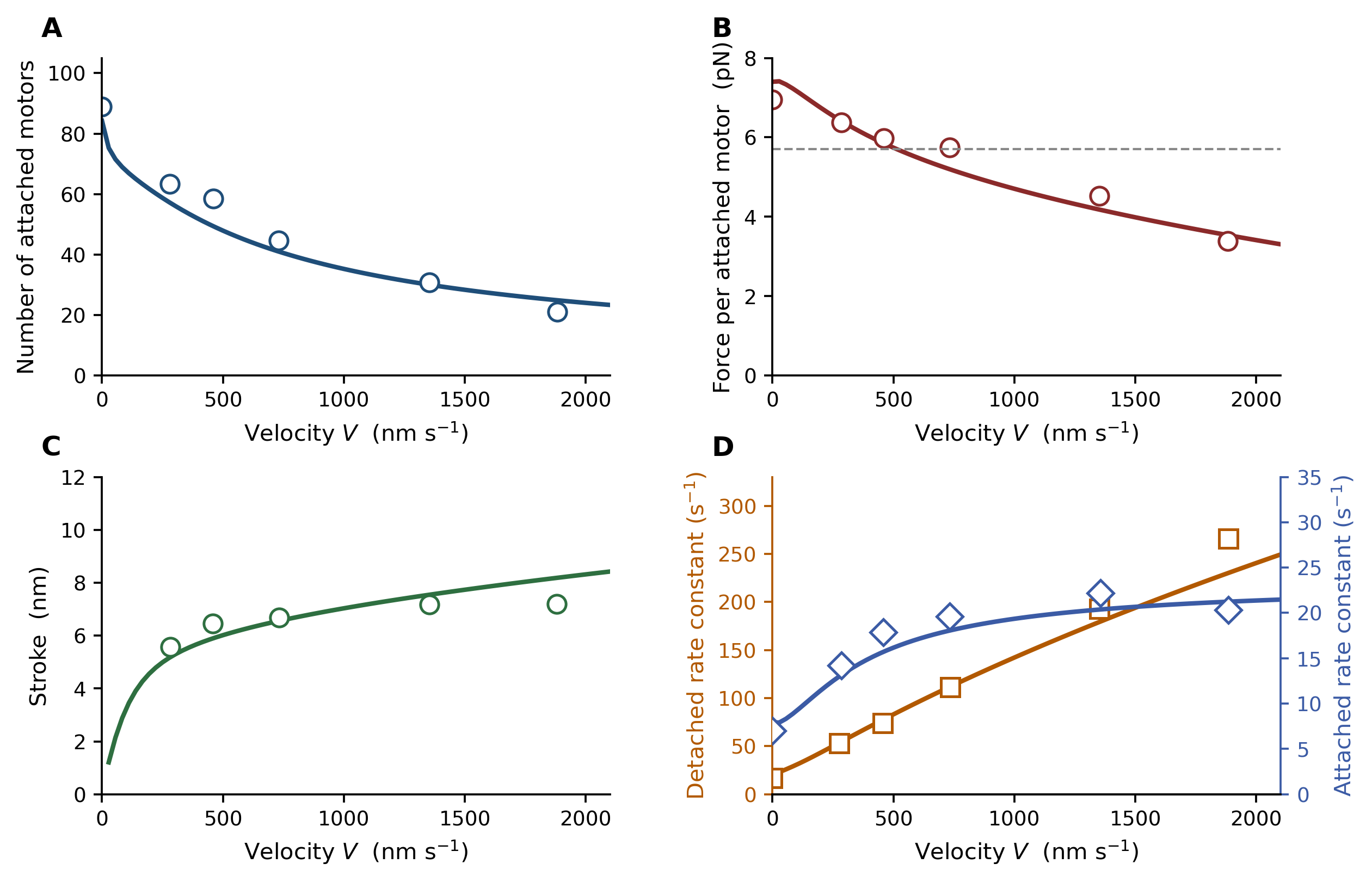}
\caption{Qualitative comparison of the reduced Huxley-type model with the
cross-bridge-level observables reported by Piazzesi et al. 
(A) Number of attached motors as a function of shortening velocity.
(B) Force per attached motor. The dashed line indicates the
isometric estimate \(5.7\,{\rm pN}\). 
(C) Apparent stroke \(L=V/g_{\rm app}\). 
(D) Apparent detachment and attachment rate constants. Symbols denote
digitized values from Piazzesi et al., while solid curves are
computed from the reduced model.}
\end{figure*}

For a given shortening velocity $V$, the steady state of the reduced model $n(x)$ satisfies
\begin{equation}\label{eq:steady_huxley_reduced}
- V \,\partial_x n(x) = f(x)\,(1-n(x)) - g(x)\,n(x), \quad x\in[-l,l].
\end{equation}
We impose the boundary condition $n(l) = 0$ when $V(t) > 0$; with this boundary condition, the ODE (\eqref{eq:steady_huxley_reduced}) has a unique solution. The macroscopic cross-bridge force is computed by integrating the mean microscopic tension:
\begin{equation}\label{eq:Fcb_reduced}
F_{\rm cb}(V)=  \textstyle \int_{-l}^{l} n(x;V)\,\partial_x G_{+}(x)\,dx,
\end{equation}
where $G_+(x)$ is the effective attached free energy induced by the fast $2\leftrightarrow 3$ equilibration, defined in \eqref{effect_G}. Although \eqref{eq:steady_huxley_reduced} can be solved analytically, the resulting expression involves nested integrals; instead, we solve it numerically using an upwind discretization.
We discretize the interval $[-l,l]$ into $N$ uniform subintervals with grid points $x_i=-l+i\Delta x$. We compute $n_i \approx n(x_i)$ from
\begin{equation}
  - V \frac{n_{i+1} - n_i}{\Delta x} = f(x_i) - (f(x_i) + g(x_i)) n_i
\end{equation}
with the boundary condition $n_{N} = 0$. We take $l = 50~\mathrm{nm}$ and $N = 10000$ in all numerical results below. This value of $l$ is used as a sufficiently large numerical cutoff rather than as an estimate of the maximum cross-bridge extension.

The model parameters involved in the reduced model are $G_i(x)$, concentrations of $\mathrm{ATP}$, $\mathrm{ADP}$ and $\mathrm{P_i}$, and the reaction rates  $\alpha_{12}(x)$ and $\alpha_{31}(x)$.
% Our goal here is qualitative. We adopt parameter values and functional forms motivated by classical cross-bridge and molecular-motor literature, but we do not attempt a systematic calibration or optimization. A data-driven parameter inference procedure, for example, fitting measured F–V curves together with ATPase rates or duty ratios, is left for future work.
We take $G_1(x) = 0$ and define
\begin{equation}\label{def_our_G}
\begin{aligned}
G_2(x) &= -\Delta G_{\rm bind} + \frac{K}{2}\, q(x)^2,\\
G_3(x) &= -\Delta G_{\rm bind}-\Delta G_{\rm stroke} + \frac{K}{2}\,q(x+d)^2
\end{aligned}
\end{equation}
where $K$ is an effective stiffness, $d$ is the stroke distance, $\Delta G_{\rm bind}$ and $\Delta G_{\rm stroke}$ encode binding and power-stroke energy changes, and $q(x) = x_0 \tanh( x / x_0)$ with $x_0 = 12~\mathrm{nm}$ is the saturating map. We treat $K$, $\Delta G_{\rm bind}$, $\Delta G_{\rm stroke} $, and $d$ as fittable model parameters. In the small-displacement regime, $q(x) \approx x$, so $G_2(x)$	and $G_3(x)$ reduce to the quadratic (Hookean) landscapes proposed in \cite{duke1999molecular}. Extending these
quadratic forms over the entire computational domain, however, produces
large energy differences and, through the detailed-balance relation,
extreme transition-rate ratios at large $|x|$. The saturating map bounds
this large-displacement behavior while preserving the quadratic form over the displacement range relevant to the simulations. The resulting landscapes are shown in Fig.~S1 in the supporting material.

% This choice reflects the difference between continuous model and MD simulations. In a continuum Fokker--Planck/Huxley-type description, the displacement $x$ is a continuous state variable and the evolution equation can advect probability mass toward the edges of the truncated $x$-domain. With a purely quadratic spring, the steady states may exhibit artificial accumulation near the imposed boundaries and the computed force can become sensitive to the chosen cutoff and boundary conditions. We therefore introduce the saturating map $q(x)=x_0\tanh(x/x_0)$ to regularize the large-$|x|$ behavior and mitigate boundary-driven artifacts, while preserving the Hookean limit in the working range. In contrast, in MD-based simulations extreme displacements are unlikely to be sampled over accessible times because geometric constraints and finite reach effectively confine the motion; as a result, the dynamics typically does not ``run into'' such artificial boundaries, and an explicit continuum-level cutoff is less critical. 

For the elementary kinetic prefactors, we use a symmetric smooth
attachment window and a strain-gated ATP-driven detachment/reset
prefactor:
\begin{equation}\label{eq:alpha12_fit}
  \alpha_{12}(x)
  =
  f_{\max}
  S\!\left(\tfrac{x+x_{\rm on}}{\sigma_{\rm on}}\right)
  S\!\left(\tfrac{x_{\rm on}-x}{\sigma_{\rm on}}\right),
\end{equation}
and
\begin{equation}\label{eq:alpha31_fit}
  \alpha_{31}(x)
  =
  g_{\min}
  +
  (g_{\max}-g_{\min})
  S\!\left(-\tfrac{x-x_{\rm det}}{\sigma_{\rm det}}\right),
\end{equation}
where \(S(z)=1/(1+e^{-z})\). The first ansatz represents a finite
reachable binding region for attachment, while the second represents
accelerated ATP-driven detachment in the negative-strain drag region.
These choices are consistent with the strain-dependent rate functions used in earlier cross-bridge models, such as the model of Piazzesi and Lombardi in Ref. \cite{piazzesi1995cross}.
The fitted kinetic parameters are
\(
  f_{\max},~~x_{\rm on},~~\sigma_{\rm on},~~g_{\min},~~ g_{\max},~~x_{\rm det},~~
  \sigma_{\rm det}.
\)
Thus the calibration is performed at the level of elementary
three-state kinetic prefactors and effective energy-landscape
parameters, rather than by prescribing arbitrary final Huxley rates
\(f(x)\) and \(g(x)\). The fitted profiles of
$\alpha_{12}(x)$ and $\alpha_{31}(x)$ are shown in Fig.~S2 in the supporting material.

%To compare with the cross-bridge-level observables reported in 
To calibrate the model using the data in \cite{piazzesi2007skeletal}, we compute the following quantities from the reduced model: the velocity dependence of the number of
attached motors, the force per attached motor, the apparent stroke,
and the apparent attachment and detachment rates. The number of
attached motors is normalized using the isometric estimate
\(
  N_{\rm att}^{0}
  = T_0 / F_{\rm per}^{0},
\)
where \(T_0=480\,{\rm pN}\) is the force borne by one half-filament and
\(F_{\rm per}^{0}=5.7\,{\rm pN}\) is the isometric force per attached
motor \cite{piazzesi2007skeletal}. This gives \(N_{\rm att}^{0}\simeq84\), about \(30\%\) of the
total number \(N_{\rm motor}=294\). For a given \(V\), the normalized attachment number is
\(N_{\rm att}(V) = N_{\rm att}^{0}\int n(x;V)\,dx \big/ \int n(x;0)\,dx\),
and the mean force per attached motor is
\(F_{\rm per}(V) = F_{\rm cb}(V) \big/ \int n(x;V)\,dx\),
where all integrals run over \([-l,l]\).
The total force is then $F(V)=N_{\rm att}(V)F_{\rm per}(V)$.
We also compute the detachment rate constant $g_{\rm app}(V)
  =
  \frac{\int_{-l}^{l} g(x)n(x;V)\,dx}
       {\int_{-l}^{l} n(x;V)\,dx},$
the apparent stroke $L_{\rm app}(V)=\frac{V}{g_{\rm app}(V)}$, and the apparent attachment rate, inferred from the apparent
two-state balance relation,
\begin{equation}
  f_{\rm app}(V)
  =
  \frac{a(V)}{1-a(V)}\,g_{\rm app}(V),
  \qquad
  a(V)=\frac{N_{\rm att}(V)}{N_{\rm motor}}.
\end{equation}
The calibration is a multi-objective problem because these quantities
are not independent. For example, the total force is
\(F=N_{\rm att}F_{\rm per}\), and the apparent rates and apparent
stroke are related through \(L_{\rm app}=V/g_{\rm app}\). We therefore
do not interpret the fitted parameters as a unique quantitative
parameter inference. Instead, the fit is used to test whether a single
thermodynamically constrained reduced model can reproduce the main
Piazzesi trends simultaneously. In the objective function, we include
the attached motor number, force per attached motor, apparent stroke,
and apparent transition rates. The resulting fitted kinetic and effective landscape parameters are
summarized in Table~\ref{tab:fitted_parameters}. This calibration is performed under a reference chemostatted nucleotide
condition, with \(c_{\rm ATP}=c_0\) and \(\mu_{\rm P_i}=0\). In the computations we set \(c_0=1\), since only the product \(c_0\alpha_{31}(x)\) enters the effective detachment rate and determines the dimensional rate scale.

\begin{table}[t]
\caption{Fitted parameters used in the reduced-model simulations. The corresponding functional forms are given in
\eqref{def_our_G}, \eqref{eq:alpha12_fit}, and \eqref{eq:alpha31_fit}.}
\label{tab:fitted_parameters}
\centering
\footnotesize
\setlength{\tabcolsep}{4pt}
\renewcommand{\arraystretch}{1.12}
\begin{tabular}{@{}lclc@{}}
\toprule
Parameter & Fitted value & Parameter & Fitted value \\
\midrule
$K$
& $1.84~\mathrm{pN\,nm^{-1}}$
& $d$
& $6.54~\mathrm{nm}$ \\

$\Delta G_{\mathrm{bind}}$
& $3.00~k_BT$
& $\Delta G_{\mathrm{stroke}}$
& $15.66~k_BT$ \\
\midrule
$f_{\max}$
& $86.89~\mathrm{s^{-1}}$
& $x_{\mathrm{on}}$
& $3.98~\mathrm{nm}$ \\

$\sigma_{\mathrm{on}}$
& $0.80~\mathrm{nm}$
& $g_{\min}$
& $7.24~\mathrm{s^{-1}}$ \\

$g_{\max}$
& $1473.14~\mathrm{s^{-1}}$
& $x_{\mathrm{det}}$
& $-8.35~\mathrm{nm}$ \\

$\sigma_{\mathrm{det}}$
& $1.50~\mathrm{nm}$
& &
\\
\bottomrule
\end{tabular}
\end{table}

The fitted landscape parameters are consistent with standard
cross-bridge scales. The displacement shift \(d=6.54\,{\rm nm}\) lies
in the nanometer working-stroke range and is close to the apparent
stroke scale reported in \cite{piazzesi2007skeletal}. The fitted stroke energy
\(\Delta G_{\rm stroke}=15.66\,k_BT\) is close to the
\(\sim 15\,k_BT\) scale used in Duke-type three-state molecular motor
models \cite{duke1999molecular,mijailovich2016three}, and the binding offset \(\Delta G_{\rm bind}=3.0\,k_BT\) is
also consistent with the binding free-energy scale used in such
models. The fitted stiffness \(K=1.84\,{\rm pN/nm}\) should be
interpreted as an effective stiffness of the reduced attached-state
landscape. It is of the same order as the compliance-based estimate
from Piazzesi et al., where the motor compliance \(c_m=0.3\,{\rm nm/pN}\)
corresponds to \(1/c_m\simeq3.33\,{\rm pN/nm}\). 
%but it is somewhat
%softer because \(K\) here represents the curvature of a coarse-grained
%effective landscape rather than the stiffness of a single isolated
%linear elastic element. The common binding offset is weakly
%identifiable in the present reduced comparison, since it enters
%\(G_2\) and \(G_3\) as a common energetic shift.

\begin{figure}[!th]
    \centering
    \includegraphics[width= \linewidth]{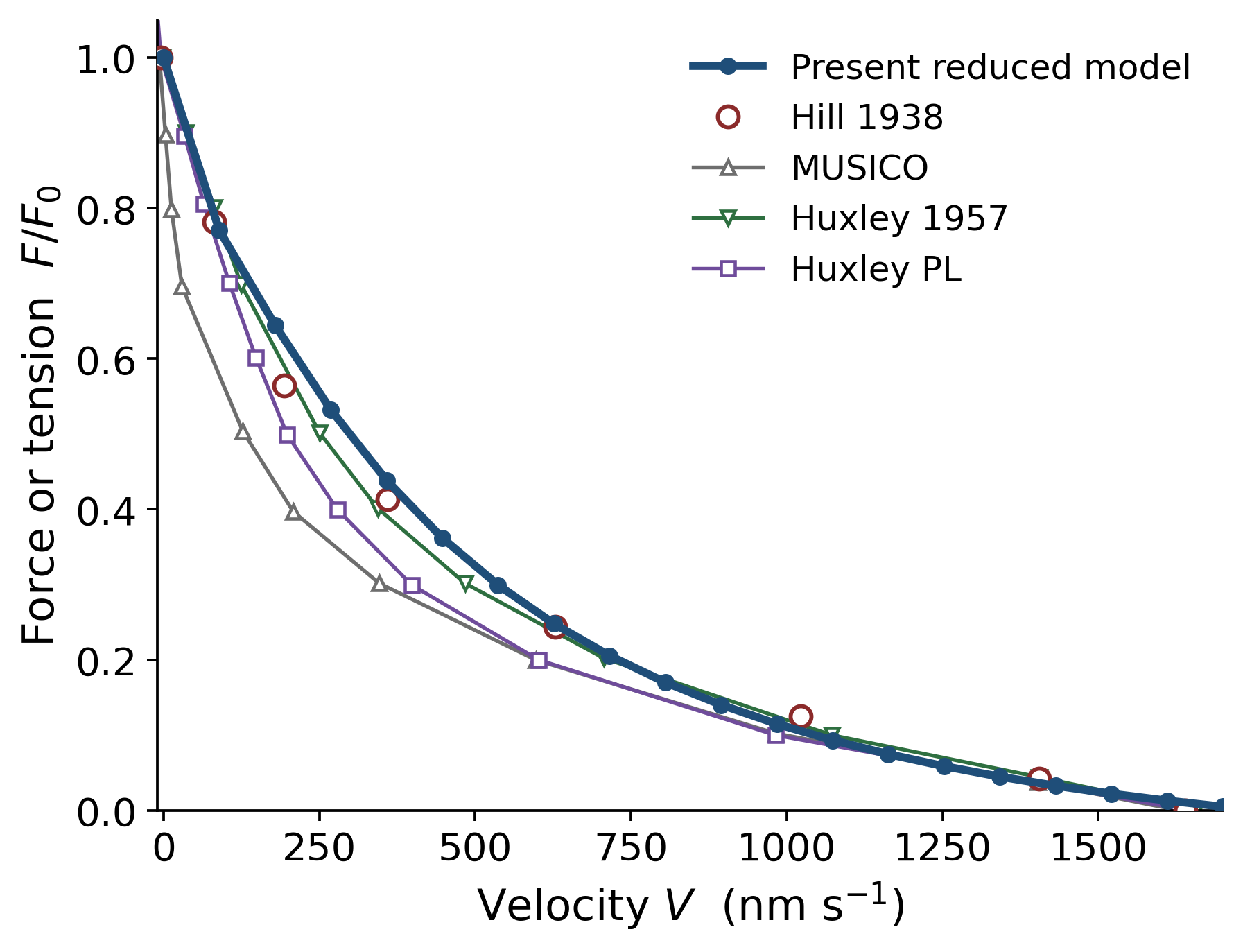}
  \caption{Force--velocity relation predicted by the calibrated reduced model. The data of A.V. Hill, and the MUSICO, Huxley 1957,
and Huxley--Piazzesi--Lombardi (Huxley PL) curves were digitized
from Fig.~7 of Mijailovich et al.~\cite{mijailovich2016three}.}\label{fig:fv}
  \end{figure}

After calibrating the reduced model against the cross-bridge-level
quantities inferred from Ref.~\cite{piazzesi2007skeletal}, we further compare the resulting force--velocity relation with the classical experimental data of
A.V.~Hill~\cite{hill1938heat} and with representative cross-bridge models, including the Huxley two-state cross-bridge
model~\cite{huxley1957muscle}, the Piazzesi--Lombardi modification \cite{piazzesi1995cross}, which enforces myosin-head conservation, and the 3D spatially
explicit MUSICO simulation of Mijailovich et al.~\cite{mijailovich2016three}. The reference data and model curves shown in Fig.~\ref{fig:fv} were digitized
from Fig.~7 of Mijailovich et al.~\cite{mijailovich2016three}. 
Since the velocity scale inferred from the Piazzesi cross-bridge-level calibration is larger than that of the classical Hill force--velocity data, we use \(c_{\rm ATP}=0.2c_0\) in Fig.~\ref{fig:fv} to reduce $V_{\rm max}$. This adjustment is consistent
with the structure of the reduced model, where ATP enters the effective detachment rate and thus controls the maximum shortening velocity. This qualitative trend has also been reported in a previous simulation study using a multistate model~\cite{chin2006mathematical}. More detailed discussion about the ATP-dependence in our reduced model can be found in the Supplementary Material. The result shows that the calibrated reduced model can reproduce the characteristic concave shape of the Hill-type force--velocity curve, and is in reasonable agreement with the classical data and models. % after setting $c_{ATP} = 0.2 c_0$ to match $V_{max}$.

Overall, these numerical results show that the thermodynamically constrained
reduced model is able to reproduce both cross-bridge-level trends and the
macroscopic Hill-type force--velocity behavior within a single set of
structured kinetic and landscape parameters. Importantly, the force--velocity
curve is not used as a direct fitting target in the calibration; it is a
derived output obtained after fitting the attached motor number, force per
attached motor, apparent stroke, and apparent transition rates. The comparison
in Fig.~\ref{fig:fv} therefore provides a consistency check that the reduced
Huxley-type limit, whose effective rates are inherited from the underlying
three-state thermodynamic cycle, remains compatible with classical muscle
mechanics. Since ATP concentration enters the effective detachment rate
through the local detailed-balance reduction, the calibrated reduced model can
also be used to explore how nucleotide conditions and energy-landscape
parameters modulate force generation and the force--velocity relation.

\section*{Conclusion}

We developed a thermodynamically consistent chemo-mechanical framework for
ATP-driven cross-bridge dynamics. Starting from a three-state Hill-type cycle,
we derived a Fokker--Planck--jump model in which ATP, ADP, and \(\mathrm{P_i}\)
are dynamical chemical variables sharing a single free-energy functional with
the cross-bridge densities, yielding a closed energy-dissipation balance that
tracks chemical input, mechanical power exchange, and irreversible dissipation.
Under chemostatted conditions and a fast-equilibration closure, the model
reduces to a two-state molecular-motor model and, in a small-diffusion limit,
to a Huxley-type transport--reaction equation
\cite{julicher1998molecular, parmeggiani1999energy, qian2000mathematical, huxley1957muscle}.
Crucially, the resulting attachment and detachment rates are not prescribed
phenomenologically: they inherit their strain, ATP, and \(\mathrm{P_i}\)
dependence from the underlying thermodynamic cycle. As a proof of concept, we calibrated the reduced model against
cross-bridge-level observables; it captures the main velocity-dependent trends
and yields a Hill-type force--velocity relation comparable to established
cross-bridge and multiscale models. This provides a basis for studying, in the
chemostatted setting, how nucleotide concentrations and energy-landscape
parameters shape force generation and the force--velocity relation.

Several next steps follow. The first is to simulate the full FP--jump model with dynamical ATP/ADP/\(\mathrm{P_i}\), which would allow us to investigate ATP-limited contraction and local nucleotide depletion beyond the chemostatted regime. A key challenge is the calibration of the additional model parameters, for which structural measurements such as X-ray diffraction patterns and microscopic information from molecular dynamics simulations may provide useful constraints \cite{mijailovich2016three, prodanovic2023using}.
Developing structure-preserving multiphysics solvers for the resulting system is an additional technical direction. Other directions include data-driven parameter inference using force--velocity, ATPase, and transient measurements \cite{wang2022some}, the explicit treatment of heat production, and coupling to macroscale mechanics \cite{borja2019functionality}.

Although we have focused on actin--myosin contraction, the same
thermodynamic principles are not limited to muscle. Similar ideas may be
useful for modeling other ATP-driven molecular motors, including
microtubule-based kinesin and dynein
\cite{peskin1995coordinated,visscher1999single,hirokawa1998kinesin,
vale2000way,atzberger2006brownian}, where chemical cycling, mechanical
motion, and energy dissipation must also be coupled consistently.

\section*{Declaration of generative AI}
During the preparation of this work, the authors used ChatGPT in order to improve the language. After using this tool/service, the authors reviewed and edited the content as needed and take full responsibility for the content of the published article.

\section*{Data Availability Statement}
This manuscript reports theoretical derivations and numerical simulations;
no new experimental data were generated. The experimental and model-comparison data shown in the figures were digitized from the cited published sources for qualitative comparison. All numerical results can be reproduced from the model equations and parameter choices in the text, and the code used to produce the figures is available from the corresponding author upon request.

\section*{Author Contributions}
Y.W.: Conceptualization, Methodology, Investigation, Writing -- Original Draft, Writing -- Review \& Editing. C.L.: Conceptualization, Methodology, Writing -- Review \& Editing.

\section*{Acknowledgments}
Y. W. was partially supported by NSF DMS-2410740. C. L. was partially supported by NSF DMS-2410742 and DMS-2216926. The authors thank Prof. Thomas C. Irving and Prof. Hong Qian for fruitful discussions.

\section*{Declaration of Interests}
The authors declare no competing interests.

% Uncomment if using bibtex (default)
\bibliography{muscle}

% Uncomment if using biblatex
% \printbibliography

\end{document}